\documentstyle[11pt,epsf,psfig,paspconf]{article}

\begin{document}

\title{Introduction to Unified Schemes}
\author{Beverley J. Wills}
\affil{McDonald Observatory \& Astronomy Department, University of
Texas at Austin, TX 78712, USA }


\begin{abstract}

The differences among apparently diverse classes of AGN are mainly the result of
viewing the central engine at different orientations, because dust, which absorbs
and scatters the light, partially covers the central source, and because synchrotron
emission is highly beamed along the relativistic jet.
Also important are factors independent of orientation: the total power
output, the unknown mover behind the eigenvector 1 relationships, and the 
radio-loudness.
These other factors may not be independent of the parameters of Unified models, such as
intrinsic jet physics, AGN dust content, and torus thickness.
We outline the basic evidence for orientation Unified Schemes, and briefly discuss their
importance for understanding the mechanisms of the central engine and its relation to
the surrounding host galaxy and beyond.

\end{abstract}


\section{Introduction}

There is a bewildering array of AGN classes.  Surprisingly, it's only in the last few years
that we've realized that much of this diversity is simply the result of viewing an axisymmetric
central structure from different angles.  Synchrotron emission from the relativistic jets is much 
brighter when viewed along the jet direction (the central engine's axis),
and a torus of dust obscures a low-latitude view of the center.
While the real physics is buried in the center, along with
clues to the enormous range of AGN luminosities, the mysterious cause of strong 
relationships among X-ray continuum and emission lines\footnote{
The so-called Eigenvector 1 or Principal Component 1 relationships link X-ray continuum slope,
linewidths and the strengths of Fe\,II emission and other lines (Wills et al. \& Francis \& Wills
contributions, this volume.)}
and the apparent radio-loud--radio-quiet dichotomy, `orientation' Unification Schemes are not
simply geometrical complications, but lead us to new ways of probing the central engine and its interaction
with the surrounding space.  This overview will emphasize QSOs, the highest luminosity AGN.
First, and most importantly, we discuss the unification of radio-loud AGN (\S2).  This is of prime
importance because the axis of the central engine can be defined by the innermost radio jets.  The
strongest arguments for unification (axisymmetry) of properties at IR through X-ray
wavebands can be made when observational properties can be referred to this axis, even for radio-weak
(radio-quiet) AGN.  In \S3, we present an idealized picture that unifies a wealth of IR through X-ray
properties, relating these to the radio axis.  Arguments in support of the general dusty torus picture
are provided by a few specific examples (\S4) and by some statistical arguments (\S5).  An overview and
speculations are given in \S6.  Additional references are given in \S7.
More detailed reviews are given by Antonucci (1993) and, for radio-loud AGN, by Urry \& Padovani (1995).
\begin{figure}
\plotone{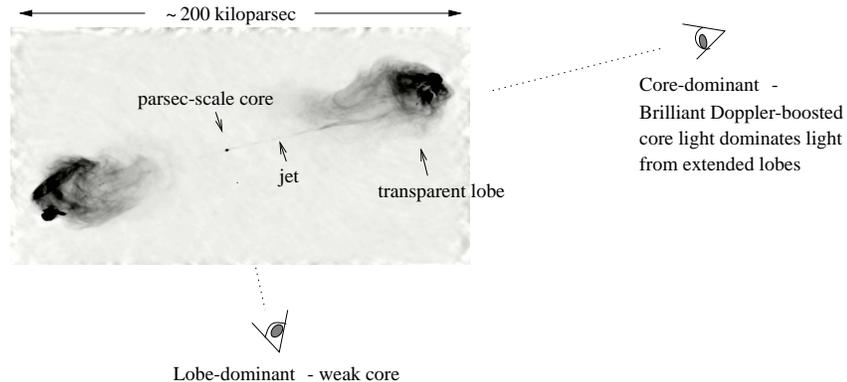}
\caption{Radio source unified scheme.  The observer below the figure views the radio source almost
perpendicular to the jet directions.  This observer sees strong, clearly separated lobes, a weak
unresolved core coincident with the optical galaxy, and faint unresolved jets linking the core to
the lobes.
The central AGN is hidden from this observer both by dust in the plane of the galaxy, and a
dusty torus whose axis lies close to the jet.  The observer to the right views the center from a 
direction close to the jet axis.  The relativistic jet beams Doppler-boosted synchrotron radiation
along the jet axis, so this observer sees a brilliant, time-variable radio core with the lobes now
seen as a relatively very faint elongated halo.  The image is a real one, actually 
that of Cygnus A, mapped at
6 cm with NRAO's Very Large Array with 0.5\arcsec resolution, viewed nearly  perpendicular to the jets.
(courtesy, Chris Carilli -- http://mamacass.ucsd.edu:8080/people/pblanco/cyga6cm\_small.gif).
}
\end{figure}

\section{Radio-loud AGN Unification}
The first recognized clues to unification came from the structure of radio sources because they can be
spatially resolved, even on parsec scales for nearby AGN.  Extragalactic radio sources often show double
lobes separated by tens to hundreds of kpc, straddling a weak compact `core' coinciding with an optical
galaxy or QSO (Fig. 1).  The emission mechanism  is synchrotron radiation. 
The lobes have steep radio spectra (F$_\nu \propto \nu^{-\alpha}$, $\alpha 
\sim 1$) so these lobe-dominated (LD) sources are most often selected by surveys at low radio 
frequencies; the cores generally have flat radio spectra, the superposition of several peaked spectra from 
optically thick sub-components.
In high-frequency radio surveys many compact, flat-spectrum sources have been discovered, often unresolved
on kpc scales (the core-dominated or CD sources), usually identified optically with QSOs, occasionally
with their almost lineless counterparts, the BL Lac objects.  The first clue to relativistically-beamed
synchrotron emission came from the intensely bright, highly variable radio cores of these BL Lac objects
and the Optically Violently Variable (OVV) QSOs,
with light travel sizes of light-hours implying impossibly high brightness temperatures -- exceeding the
Compton scattering limit.  The second clue
came from Very Long Baseline Interferometry (VLBI) that probed the compact radio sources on scales of
milliarcseconds.  Blobs were ejected, apparently at superluminal speeds in the plane of the sky.
As blobs travel towards the observer at relativistic speeds, $\ga$0.7c, the emitted signal has less far to
travel, that is, the blobs tend to `catch up' with their forward-emitted
radiation; thus, for a distant observer looking close to the direction of ejection, events appear to
happen closer together in time, and the blob appears to travel at speeds
up to 10c or more.  Under these conditions, the synchrotron radiation will be forward-boosted into an angle 
$\gamma^{-1}$ rad,\footnote{
The Lorentz factor
$\gamma = (1 - \beta^2)^{-1/2}$, and
the Doppler factor $\delta = \gamma^{-1} (1 - \beta {\rm cos}\theta)^{-1}$.
} 
and with amplification or Doppler boosting up to a factor $\delta^{n+\alpha}$ 
(n = 2 -- 3), simultaneously
explaining the brilliant compact sources.
The most general Unification Scheme for radio sources states that all CD
sources are LD sources viewed close to the jet axis (within 15\deg--20\deg\ of the line of sight.)
This is illustrated in Fig. 1.

For a random distribution of orientation angles one expects to see $\sim2\gamma^2$ LD sources  for
every CD source.  This is more-or-less consistent with the statistics of radio sources, taking
into account the frequency and flux-density limits of radio surveys (Urry \& Padovani 1995).  This is nice,
but somewhat unexpected in the sense that it is not obvious why 
the electrons producing the
Doppler-boosted emission should have
same $\gamma$ as the flow pattern determined from apparent superluminal motion: first,
the boosted radiation is dominated by those electrons that happen to be traveling in the observer's
direction, so a spread in direction of the electron stream could result in
an actual beam width much broader than ${\gamma}^{-1}$ (Lind \& Blandford 1985).  
Second, the pattern speed is probably an apparent speed, representing the phase velocity of shock
fronts in the jet.
Several 
observations suggest that the pattern speed is in fact close to the speed of the Doppler-boosting electrons,
so we're in luck.

The optically-thin emission from the lobes is isotropic, so
what this means is that we have a measure of orientation of the central engine (or inner jet), the core 
dominance, defined by: 
\vskip 1mm
 R = observed core flux-density/unbeamed lobe flux-density
\vskip 1mm
\noindent converted to rest-frame frequencies, often chosen as 5 GHz.

This Unified Scheme predicted the following:\\
$\bullet$ The jets should have very different brightnesses.  The approaching jet should be brighter than
the receding jet, by
$[(1+\beta {\rm cos}\theta)/(1-\beta {\rm cos}\theta)]^{n+\alpha}$.  In fact the fainter jet is
rather faint even when it's thought that the jets are close to the sky plane.  These jet
asymmetries are observed out to several Kpc (in powerful 3CR sources, at least).  The lobes are
of similar brightnesses to each other because by the time the jet has reached them it has become much less
relativistic, and environmental factors can dominate (Bridle et al. 1994; Dennett-Thorpe et al. 1997;
Wardle \& Aaron 1997).\\
$\bullet$ The brighter jet is on the near side of the nucleus.
This is beautifully demonstrated by radio Faraday
depolarization which is almost always greater for lobe emission on the side of the fainter (invisible) jet.
Recent observations
have shown that the depolarization occurs in a foreground screen associated with the 
radio source, rather than being intrinsic to the lobes, so overcoming an earlier
objection to this argument (e.g., Leahy et al. 1997; Morganti et al. 1997).\\
$\bullet$ Slight curvature of jets, when foreshortened as expected in CD sources, should appear to be
much larger than in LD sources.  This is observed (Hough \& Readhead 1989; Cohen \& Unwin 1982.)\\
$\bullet$ Core-dominated sources, whose apparent brightness often exceeds the Compton limit,
should be seen as
strong emitters of X-rays and $\gamma$-rays, a prediction strikingly confirmed (Urry 1999).\\
$\bullet$ Apparent superluminal speeds should decrease with decreasing core dominance, and this is
observed (Browne 1987; Vermeulen \& Cohen 1994).\\
$\bullet$ The CD sources have relatively faint diffuse halos with structure,
often double, and luminosity consistent with their being the extended doubles seen end-on.  On this basis 
Antonucci \& Ulvestad (1985)
used the converse to argue that the radio source Unified Scheme must be basically correct:  CD
sources seen at larger angles to the jet must be seen as powerful steep spectrum radio sources dominated
by extended, basically double-structured emission, and can account for most of the fraction of observed
LD sources.\\
$\bullet$ There is an inverse relationship between R and projected linear size.  Barthel (1989)
showed that the projected (apparent)
radio linear sizes for 3CRR QSOs are smaller than for the lower-R FR II radio galaxies, as predicted.  A
quantitative demonstration of this relationship has been thwarted by selection effects and probable 
linear size evolution (Gopal-Krishna et al. 1996; Neeser et al. 1995).

While many radio sources show clear double-lobed structure (e.g., Fig. 1), in real life structure can
be more complex on both parsec and kiloparsec scales.  About 20\% of sources in low radio-frequency
surveys consist
of intrinsically compact double or single sources -- Gigahertz Peak Spectrum (GPS)
and Compact Steep Spectrum (CSS) sources (Urry \& Padovani 1995).  It is not yet
clear whether unification works for these intrinsically compact sources.
Also, the LD sources come in two flavors, FR\,I and FR\,II (Fanaroff \& Riley 1974;
Owen et al. 1996).
The lobes of the more luminous\footnote{
$L_\nu({\rm 178 MHz}) \ga 1.3 \times 10^{33}$ erg s$^{-1}$ Hz$^{-1}$, for H$_o = 50$ km s$^{-1}$ Mpc$^{-1}$.
} FR\,II sources contain bright hotspots and are edge-brightened where they ram into the 
surrounding medium, whereas the less luminous FR\,I lobes are relaxed and show neither of these features.

\section{Unification of the Optical, Infrared and X-ray Properties}

The central regions of luminous AGN are essentially unresolved at non-radio wavelengths,
and their classification is based on optical spectra.

Optical spectra of Type 1 are defined by their broad emission lines ($\sim 6000$ km/s FWHM),
which originate in gas within 1 pc of a central supermassive black hole.  They may also show
narrow lines that arise from low density gas on scales of parsecs to kiloparsecs.  Type 1
spectra are accompanied by a broad optical-EUV bump attributed to thermal emission with a range
of temperatures, 10$^4$ K to 10$^5$ K, arising from an accretion disk that feeds the
black hole (Shields 1978; Malkan \& Sargeant 1983; Mushotzsky 1997).
These observed properties define the QSOs and their
lower-luminosity cousins, the Seyfert 1 nuclei.  Optical surveys for these objects usually
depend on selection by blue or UV colors and strong broad emission lines.
In radio-loud AGNs it appears that Type 1 spectra are found in FR\,II radio sources, but almost never
in the lower luminosity FR Is (Falcke et al. 1995).

Optical spectra of Type 2 are those seen to have narrow lines only, with a spectrum and
velocity dispersion indistinguishable from the narrow lines of Type 1 spectra.
The optical continuum is dominated by starlight, and there is no convincing evidence for a
optical-EUV bump associated with narrow line spectra.

When seen in the abscence of obscuration, the thermal Type 1 and Type 2 emission lines and
optical-EUV bump are essentially unpolarized.

In core-dominant radio-loud AGNs  a steep-spectrum, polarized, smooth spectral energy distribution
of an IR-optical-UV
synchrotron component can contribute (the blazars), in some cases overwhelming the emission 
lines and accretion continuum of a Type 1 spectrum.  The spectra of BL Lac objects often
appear featureless, and are always
dominated by this synchrotron continuum, and may have no, or intrinsically only very weak broad lines.
Narrow emission and absorption lines may be present, as for any AGN.

Fig. 2 illustrates how AGN can appear drastically different at different orientations.
It shows (not to scale) a cross section through the dusty torus, whose shadow is shown in gray.
A radio jet axis is shown aligned with the axis of the dusty torus.
The broad lines arise from dense gas of the broad-line region (BLR) and narrow emission lines
from less-dense NLR gas.  These are shown as filled and open circles in Fig. 2.
Refer to Fig. 2 in the remainder of this section.

\subsection{Powerful FR\,II Radio-Loud (RL) AGNs.}

\begin{figure}
\plotone{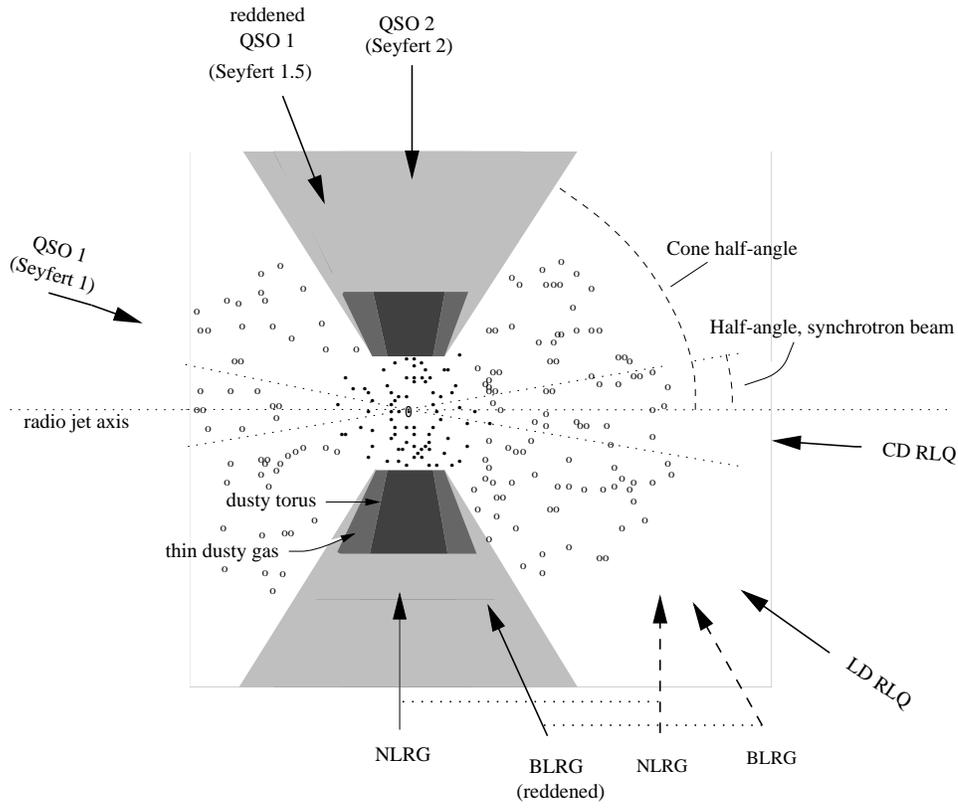}
\caption{A similar central engine viewed from different inclinations to the axis gives
rise to different classes of AGN.  Dense, high speed ($\sim 7000$\,km/s) gas of the BLR
(solid dots) may be exposed to view, partially absorbed, or completely hidden by the dusty
torus (shadow shown in grey).  The dashed arrows show directions from which a polarized,
scattered light spectrum of the center
can arise.  At least half of the lower-velocity NLR gas (open circles) is seen from any view,
except the highest ionization NLR close to the center.
Imagine you, the observer, are situated at very great distances so that, e.g., the two
views of a NLRG are seen in parallel light, as indicated.
Left and top of the diagram are radio-quiet (RQ) classifications, and right and bottom are
radio-loud (RL) classifications. 
}
\end{figure}

One can see the BLR and accretion continuum directly from within the cone whose edge is
defined by the shadow
of the torus.  Within 15\deg\ to 25\deg\ of the jet axis the QSOs are CD with R$\ga1$,
and the
strong variable synchrotron emission is revealed by a steep polarized infrared-UV continuum
that may swamp the unpolarized blue bump and emission lines.  These QSOs are therefore often 
seen as optically violent variable (OVV) or highly-polarized (variable) QSOs (HPQs).
Beamed Compton-scattered radiation from the jet contributes to and flattens the X-ray spectrum.

At larger angles to the axis the QSOs are LD (R$\la1$), and the beamed synchrotron
emission is no longer detectable against the optical-EUV bump.
The broad H$\beta$ line and broad base of the C\,IV$\lambda$1549
emission lines are broader than in CD QSOs, perhaps indicating a disk-like configuration for the
inner BLR.

At even larger inclinations of the jet axis, perhaps 40\deg\ to 50\deg, R decreases further, with
increasing projected lobe separations (Barthel 1989).  Broad lines may be very broad,
up to 10,000 or even 20,000 km s$^{-1}$ in Doppler width (FWHM), and are usually reddened and weaker
relative
to the NLR emission (Grandi \& Osterbrock 1978; Hill, Goodrich \& DePoy 1996).  Some of these 
reddened QSOs and broad-lined radio galaxies (BLRG) show weak, polarized broad-line and 
continuum spectra ---
evidence for a scattered Type I spectrum that is combined with the reddened view through the 
thinner regions of the torus.

At even larger inclinations emission from the NLR completely dominates the spectrum (Fig. 3).
These are the narrow-lined radio galaxies (NLRG) with their Type 2 spectra.  Direct light
from the BLR and optical-EUV continuum may be completely obscured, and even the higher-ionization 
NLR emission ([O\,III]$\lambda$5007) may be partially obscured by the dusty torus.  Starlight completely 
dominates the near IR-optical continuum.  The inner regions
may be seen indirectly however, as a faint polarized spectrum that has been scattered from within the
opening angle of the torus -- the `ionization cone' or `scattering cone'.  Even the scattered
light may be absorbed as it grazes the torus -- or passes through the host galaxy.
These ionization cones are sometimes resolved by HST and ground-based imaging.

The ``3$\mu$m Bump'' arises from warm dust near evaporation temperature ($\sim1700$ K).
This equilibrium temperature determines the distance of the inner torus from the central
heat source, $\sim L_{46}^{1/2}$ pc for luminous QSOs of bolometric luminosity
$L_{46} \times 10^{46}$ ergs s$^{-1}$.  This is present in most, if not all, RL QSOs (RLQs).

\begin{figure}
\plotone{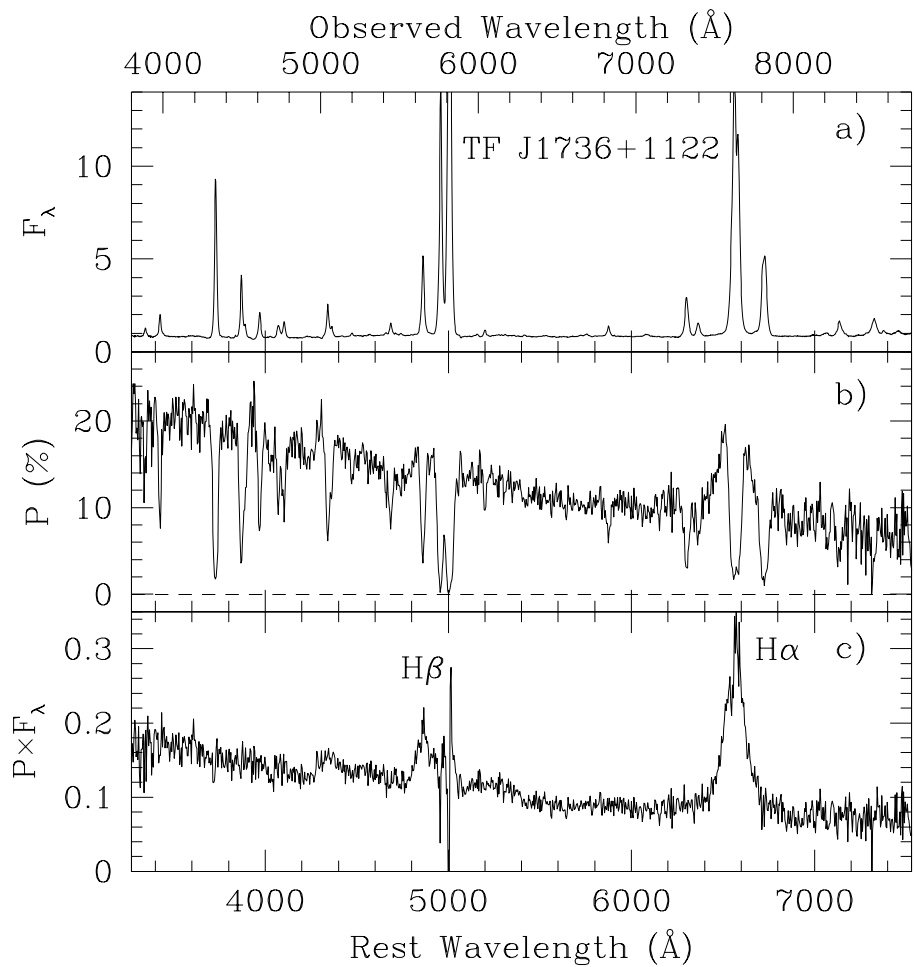}
\caption{Spectropolarimetry of the QSO 2 TF J1736+1122.  The upper panel shows the spectrum in
total (polarized plus unpolarized) light, dominated by the , strong NLR
emission characteristic
of a Type 2 spectrum.  The spectrum in the middle panel shows the degree of polarization.
Note how the degree of polarization drops at the wavelengths of the strong, essentially unpolarized,
QSO 2 narrow lines.  
The lower panel shows how the faint scattered-light QSO 1 spectrum is revealed in polarized light.
Note the broad, blended Fe\,II emission peaking near 4570\AA\ and 5250\AA, and the broad Balmer
lines in the polarized flux spectrum.  
(Figure from Tran et al. 1999). 
}
\end{figure}

\subsection{FR\,1 AGN and BL Lac Objects}
Does the above Unified Scheme also hold for the FR\,I AGN, which are less luminous?
Most FR\,I AGN are LD sources and appear optically as radio galaxies.
No FR\,I radio galaxy had shown anything but weak narrow line emission until recently,
when Lara et al. (1998)
discovered an FR\,I nucleus with broad emission lines.  
The BL Lacs, which are always CD, are probably the LD FR\,I radio galaxies seen at small
angles to the radio jet.  Consistent with this, is their lack of, or only very weak,
broad emission lines.  The extended radio fuzz around BL Lac 
cores is consistent in luminosity and structure, and probably in size as well, with the
extended radio emission of double-lobed FR\,I radio galaxies.   The evidence for relativistic jets
in BL Lac objects has already been mentioned: (i) from their high apparent brightness temperature
and rapid variability.  Their radio-IR-optical-UV continua are smoothly connected, linearly 
polarized and highly variable, as expected for a beamed synchrotron origin.  The BL Lacs are so far
the only AGN detected at TeV ($\gamma$-ray) photon energies -- attributable to beamed radiation from
the relativistic jet, and their lower redshifts (smaller distances) than most core-dominant RLQs.  
(ii) The jets are also 
seen in VLBI images.  While the extended lobes of FR\,I radio galaxies are relaxed, with no edge
brightening or hotspots, their inner radio jets are indistinguishable from those of FR\,II sources,
and apparent superluminal motion has been measured (Giovannini et al. 1998).

\subsection{Radio Quiet (RQ) AGN}

Radio-quiet (RQ) AGN include both the RQ QSOs (RQQs), which are of course optically luminous,
and the Seyfert galaxies, which are less luminous\footnote{
We reserve the term `Seyfert' for low-luminosity radio-quiet AGN.
Spectroscopically, the Seyfert 2 AGN are the radio-quiet equivalents of the low-radio-power NLRG.
There appear to be no high radio-power equivalents of the Seyfert class.  Apparently the
production of FR\,II radio sources requires the presence of a powerful optical central engine,
as indicated by a QSO.
}.  Their
radio emission is weak, though detectable.  

For nearby Seyfert 1s and 2s, ionization-scattering cones can be related to their radio jet
axes, and their unification by
orientation is quite well established both statistically, and for many individual Seyfert 2s
where a buried Seyfert 1 has been revealed in polarized light (\S4 \& \S5).  This does not 
preclude the existence of a small fraction of genuine Seyfert 2s where the BLR is intrinsically weak, 
perhaps temporarily (e.g., N4151, Penston \& Perez 1984).

Does this unified scheme hold for radio-quiet QSOs?
Kukula et al. (1998) argue that the radio-quiet QSOs form the high radio and optical luminosity 
extension of the Seyfert 1 luminosity functions, which is suggestive that it does.  While it has
been shown that there is patchy
covering by absorbing and scattering dust in radio-quiet QSOs (\S6), the dust geometry has not been 
clearly related to the axis of the central engine.  Also, no large, unbiased survey has been made for
QSO 2s -- whatever they may be expected to look like.

\subsubsection{Radio-quiet QSOs \& the Axis of the Central Engine}

The QSO radio luminosity function appears bimodal (Hooper et al. 1995) so a distinction has been made 
between radio-loud QSOs (RLQ) and radio-quiet QSOs (RQQ).  A somewhat arbitrary definition often used 
to discriminate RQQ from RLQ is that RQQ have a rest-frame flux-density ratio 
F$_{\nu}$(5 GHz)/F$_{\nu}$(B-band)
$\la 10$.  Results from deep radio surveys ({\it FIRST}, White et al. 1999 preprint) will show whether RQQ
and RLQ are really separate populations;  the lower the survey frequency the better, to avoid the
complications of beamed radio emission.
Recent surveys of the radio properties of radio-quiet QSOs (e.g., Falcke, Wilson, \& Ho 1997;
Kukula et al. 1998;
Blundell \& Beasley 1998a; Kellermann et al. 1994, 1998) show strong evidence for jet-producing central 
engines in many.  Falcke et al. (1996a, b) make a strong case that a few QSOs with radio flux densities
intermediate between radio quiet and radio loud are naturally 
explained as Doppler-boosted radio-quiet QSOs, the radio-quiet equivalent of CD RLQs.  For these they
derived similar $\gamma$(Doppler boosting).  This is supported by the discovery of superluminal motion
in at least one radio-quiet QSO (Blundell \& Beasley 1998b).
These new results are 
important for the interpretation of optical properties in terms of
an axisymmetric central engine.  However, it must be appreciated that only the least radio quiet of the
radio-quiet QSOs can be investigated for jet structure and superluminal motion.

\subsubsection{Broad Absorption Line QSOs}

An important subclass of QSOs are the broad absorption line QSOs (BAL QSOs).  These objects show broad
absorption features, blueshifted from a few thousand km s$^{-1}$ up to 45,000 km s$^{-1}$ or even more,
from the systemic redshift.  The observed radial terminal velocity of these BALs is strongly 
anticorrelated with radio power (Weymann 1997). 
The soft X-rays are completely absorbed in these objects (Green \& Mathur 1996).  These absorption features
provide evidence for high-ionization and dusty, low-ionization nuclear outflows (see \S4 below).
Three main results have led to the idea that all RQQs have BAL outflows:\\
(i) In individual BAL QSOs, from the limits on scattered light in the deep troughs, one can deduce
that the absorbing gas covers  $\sim$10\% of the central continuum.\\
(ii) The BAL QSOs and the non-BAL QSOs have very similar UV emission line spectra (Weymann et al. 1991).\\
(iii) BALs are observed in $\sim$10\% of all RQQs.\\
Can this `orientation unification scheme' be related to the above scheme for radio-loud QSOs?
Unfortunately, most radio-quiet BAL QSOs have only weak, unresolved radio emission, too weak for
present observations at milliarcsecond (VLBI) resolutions.  So no jet axis can be determined.
Nevertheless, the
radio structure of radio-loud BAL QSOs may soon provide important clues to orientation and 
radio-loud--radio-quiet differences.  There may be other clues to axisymmetry and inclination, however.
The BAL QSOs were the first objects as a class to show significant optical linear polarization -- not
counting the highly variable blazars whose high polarization arises from beamed synchrotron-emitting jets.
 Spectropolarimetry of several BAL QSOs shows evidence for reddened direct views of the
nucleus and multiple scattered light paths.  Unlike most other `buried' Type 1 objects, the broad
emission line polarization is generally low compared with the continuum, suggesting scatterers mixed with
or at distances less than the BLR.  In principle, the polarization position angle, and modeling the 
degree of polarization, may lead to further clues to the radio-quiet QSOs' inner structure (see the
discussion of IRAS 07598+6508 \S4.7, Fig. 4).

Serious doubt about the above simple orientation unification of BAL QSOs and non-BAL QSOs is
suggested by the probable association of strong Fe\,II emission and weak [O\,III]\,$\lambda$5007
(NLR) emission with the existence of BALs, that is, of `Principal Component 1' with the existence
of BALs (Boroson \& Meyers 1992; Turnshek et al. 1997).  Several investigations suggest a range in
covering factor for BAL material, so the truth may lie in orientation, together with a range in
covering factor.

\begin{figure}
\plotone{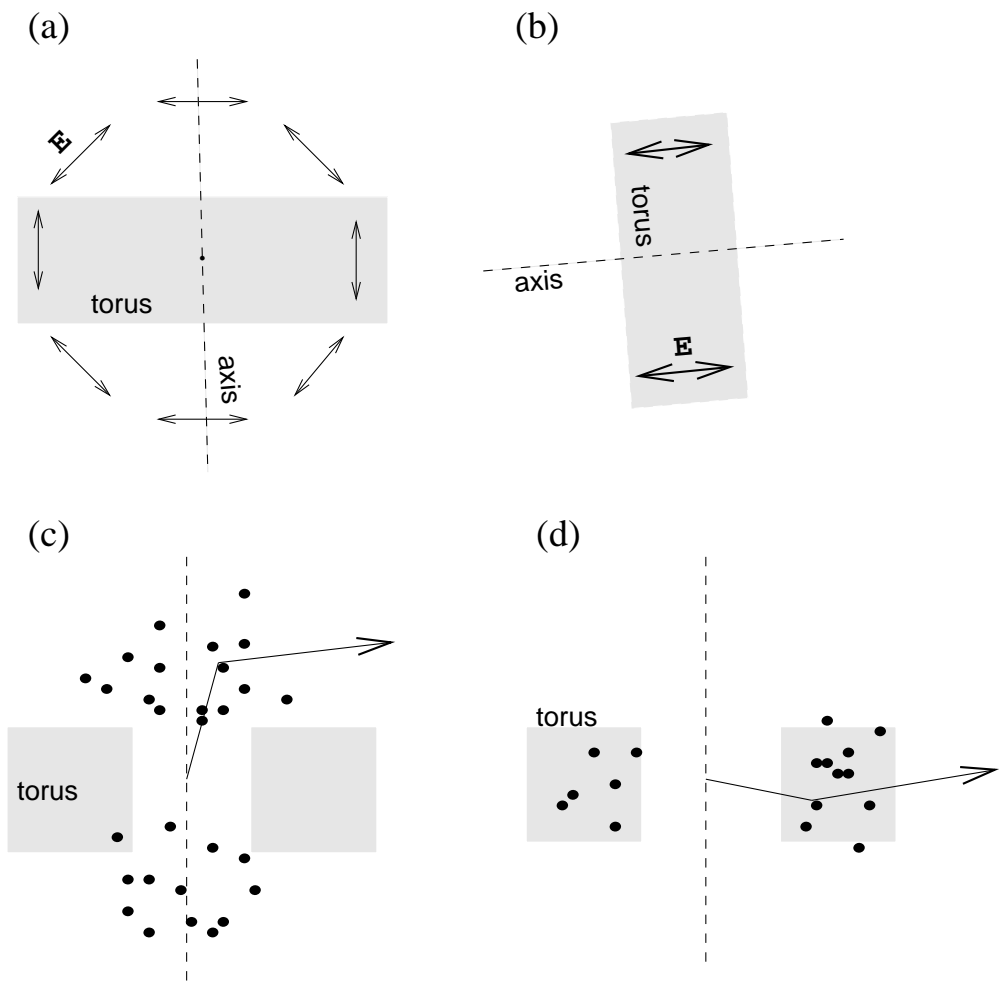}
\caption{(a) Observer's view: Scatterers above or below optically-thick torus
produce polarization {\bf E}-vector perpendicular to the axis. (b) Observer's view:
Scattering by material in an optically thin region produces {\bf E}-vector parallel to
the axis.  (c) Lateral view of light path, case (a).  (d) Lateral view of light path,
case (b).  }
\end{figure}

\section{Type 1 AGN Revealed by Scattered Light: Case Studies}
By now there are many examples, so
we choose a few for illustration, and give references to some others.
\subsection{NGC\,1068, a Radio-Quiet Hidden Seyfert 1 Galaxy}
 NGC\,1068 is the prototypical Seyfert 2 galaxy and the prototype of Type 1 nuclei buried within a
dusty torus.  The following situation is similar to that illustrated  in Fig. 3.
Spectropolarimetry revealed a faint polarized Type 1 spectrum (Miller \& Antonucci 1983).
In total light the
scattered Type 1 spectrum is dominated by an essentially unpolarized Type 2 spectrum, nearly 100 
times stronger.  The Type 1 spectrum shows wavelength-independent polarization from optical to UV
wavelengths ($\sim$16\%), indicating electron scattering, and the polarization is perpendicular to 
the weak radio jet (Fig. 4(a) \& (c)).  Direct light from the center (IR to X-rays) is obscured 
(A$_V \ga 25^{\rm m}$) by the edge-on torus, whose outer extensions have been
imaged in the near infrared (on a scale of tens of pc), in thermal emission from ionized clouds on 
VLBA scales of
$\sim$1 pc, and detected in H$_2$O and OH maser emission.  Even the unresolved radio nucleus is partially
absorbed.  The ratio of scattered optical continuum to X-ray continuum is like that for unobscured Type 1
nuclei, indicating electron scattering of the X-ray spectrum as well.  High equivalent width
Fe-K$\alpha$ near 6.4 keV from the unobscured scattering region was predicted and confirmed.
At optical and UV wavelengths, a one-sided `cone' of high-ionization and scattering gas, formed by
the shadow of the dusty torus,
subtends an opening angle of about 40\deg\ projected on the sky plane (Pogge 1988).
A bicone is detected in the near infrared (Packham et al. 1997).
As in many other such obscured Seyfert 1s, the ionizing photon flux seen by the emission-line gas is 
significantly greater than can be accounted for by our line-of-sight view of the nucleus (Wilson 1996).
In fact, this was one of the first clues that the central continuum was anisotropic.
In addition to the electron scattered view of the center of NGC 1068, Miller, Goodrich, \& Mathews (1991) 
discovered a second fainter view in scattered light from a dust patch $\sim5$\arcsec NE of the nucleus.
Here, a polarized Seyfert 1 flux spectrum rising to short wavelengths 
indicated Rayleigh-type scattering from dust grains.  
The dust-scattered H$\beta$ profile from the second position shows a `Narrow-lined Seyfert 1' spectrum,
suggesting that the electron-scattered profile from the first position is broadened by thermal motion
of the hot electrons.

\subsection{NGC\,4258, the maser galaxy}
 While NGC\,4258 is only a very low-level AGN, it has special importance because of its closeness
($\sim$7 Mpc) allowing good linear resolution, but especially because it contains a thin edge-on
warped molecular disk, enabling the exciting discovery of water vapor maser emission.  The 
radial velocity curve of the maser lines can be mapped at VLBI resolution, and shows a very 
accurate r$^{-2}$ dependence to within 0.13 pc of the center (Miyoshi et al. 1995).
This led to an estimate of $\sim 4 \times 10^7$ M$_{\sun}$\ 
for the central mass, and the best evidence yet for a supermassive black hole in an active galaxy.  
Spectropolarimetry revealed a faint AGN-like blue continuum and emission lines polarized at 5--10\%.  
The emission lines were broad for a low luminosity nucleus ($\sim$1000 km s$^{-1}$).  
Polarization is perpendicular to the disk axis.  These observations suggest that the molecular disk 
or torus obscures an AGN central engine, albeit a low-power one.  High-resolution imaging in the near
IR actually reveals an unresolved nucleus, suggesting A$_V \sim$17$^{\rm m}$ (Chary \& Becklin 1997).
A few additional maser sources with similar properties have since been discovered (e.g., Trotter et al.
1998).

\subsection{IRAS 09104+4109, a Hidden Radio Loud FR\,I? QSO}
This IRAS-discovered AGN is identified with a very luminous cD galaxy (z$\sim$0.4) in a rich cooling-flow 
cluster.  It is important because its nucleus is of QSO luminosity ($\sim 10^{12.5} h^{-2}$ L$_{\sun}$)
and excitation, with a powerful Type 2 spectrum -- a QSO\,2.  A scattered QSO Type 1
spectrum is seen in polarized flux (p$\sim$20\%).  The radio structure is more like FR\,I than FR\,II,
which makes this object of importance for both FR\,I and FR\,II unified schemes.  
It is one of the few `hidden' QSOs to show both opposed scattering-ionization cones (opening angle
$\sim46$\deg).  These giant cones extend $\sim5$\,Kpc from the nucleus.

\subsection{OI\,287 = Q\,0752+258, a LD radio-loud QSO with a thin torus}
This is a LD radio-loud QSO with high, wavelength-independent degree of polarization, 
indicating an electron-scattered Type 1 spectrum and a completely obscured direct view of the nucleus.
Unpolarized narrow lines of high equivalent width suggest a direct view of the NLR and an obscured
continuum.
The wavelength-independent polarization parallel to the radio jets indicates scattering in an
optically-thin edge-on disk (Fig. 4(b) and (d)).  While OI\,287 is of high radio power, the extended radio
features appear more like bridges than jets;  in this respect it is more like IRAS\,09104+4109, which has
been suggested to have FR\,I radio structure.

\subsection{3CR\,68.1, the Most Luminous 3CRR QSO}
This z=1.2 AGN is special because it is the most luminous, most lobe-dominant QSO in the 3CRR catalog.
Its optical-UV
continuum is the reddest known -- and it shows a highly polarized Type 1 spectrum, with degree of
polarization increasing to $\sim$10\%
in the UV.  This is interpreted in terms of a highly reddened direct spectrum, combined with a reddened 
scattered spectrum.  Like OI\,287, 3CR\,68.1  shows strong associated absorption, which
can be understood as the result 
of light from the center passing through absorbing clouds associated with
the thinner regions of the torus (Fig. 2).

\subsection{IRAS\,13349+2438, a RQQ}
This is the archetype of highly-polarized radio-quiet QSOs.  It was the first QSO to be discovered in
the mid infrared (12\micron -- 100\micron) by IRAS.
In the optical-UV it shows a strong, but reddened, Type 1 spectrum (z = 0.10), with extremely weak NLR 
emission.  Its continuum and broad lines are increasingly
polarized towards the UV, up to 10\%, and its polarization ({\bf E}-vector) is aligned with the major axis
of a galaxy disk
visible in the near IR.  The explanation is in terms of a combination of a direct spectrum partially 
obscured by a dusty torus,
and a much less-reddened (polarized) scattered spectrum.  X-ray variability suggests a direct view to the
center, but surprisingly, in view of the optical reddening, IRAS 13349+2438 is strong in soft X-rays
($0.2$keV -- $0.6$keV) with
no sign of the expected cold absorption.  This was one of the first examples of X-ray `warm absorption'.
The dust appears to be mixed with warm ionized gas that produces absorption only at higher photon energies,
e.g., O\,VII and O\,VIII edges at $\sim$0.7 keV.

\subsection{IRAS\,07598+6508, a BAL QSO}
Another IRAS-discovered radio-quiet QSO, IRAS\,07598+6508, at z=0.15, was found first to have high
polarization, 
leading to the discovery of extremely strong blue-shifted broad absorption lines (BALs) in the UV.
Spectropolarimetry showed a polarized continuum and unpolarized broad emission lines, unlike the
QSO 2s and Seyfert 2s investigated so far.  This was the first object
for which an increased degree of polarization was seen in an absorption line -- indicating {\it 
decreased} dilution of a polarized continuum by an unpolarized absorbed continuum component.  Because the 
broad emission lines are apparently not scattered, but part of the continuum is, the scatterers must 
lie nearer the QSO continuum source than the BLR, and be distinct from the absorption region.  This is 
an important constraint on scattering geometry.
Like other BAL QSOs it is not detected in X-rays, suggesting complete absorption by neutral gas with
N$_H>10^{23}$ cm$^{-2}$.
IRAS\,07598+6508 is also a `super Fe\,II' emitter.

\begin{figure}
\plotone{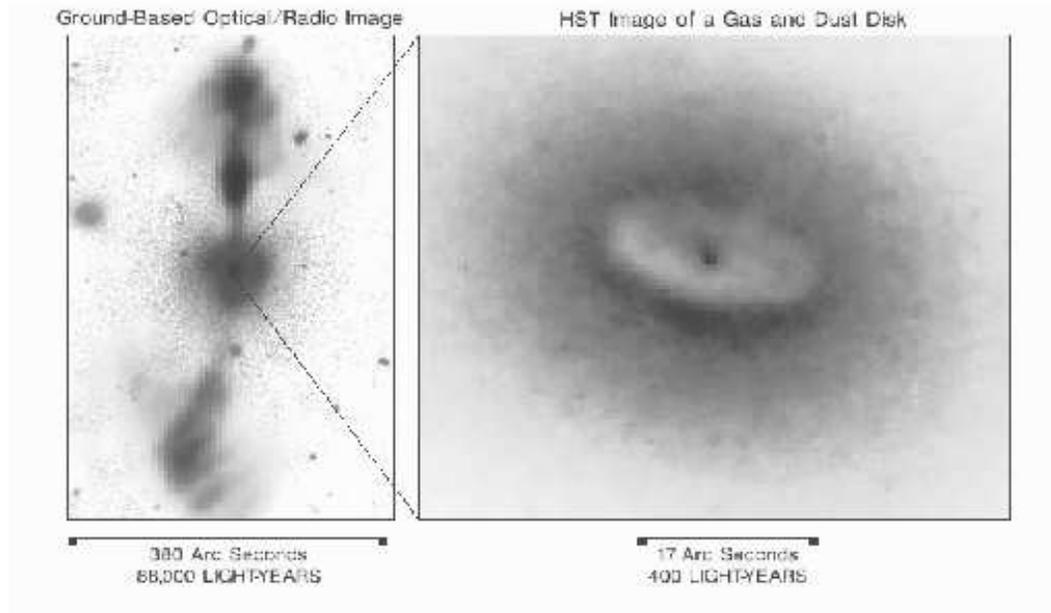}
\caption{Left: The classic FR\,I radio source, 3C\,270, showing the radio nucleus and $\sim$16 Kpc-long
jets superposed
on a ground-based optical image of the associated galaxy, NGC\,4261.  Right: The HST optical image of
the nuclear regions of NGC\,4261, showing the increasing stellar surface brightness towards the
bright, unresolved nucleus, blocked by a dusty disk (shown lighter).  The jets are almost aligned with
the disk axis.
NGC\,4261 is a LINER galaxy (Low Ionization Nuclear Emission-line Region), with dynamical evidence 
suggesting a supermassive black hole (Ferrarese et al. 1996).
Image from http://oposite.stsci.edu/pubinfo/Old.html.  Observation by Jaffe et al. (1993).  For updated
optical images, see Jaffe et al. 1996, and especially Ferrarese et al. 1996).
See also the NGC 7052 dust disk at http://oposite.stsci.edu/pubinfo/pr/1998/22/
}
\end{figure}

\section{Dusty Torus Geometry:  Statistical Relationships}
\subsection{Evidence for Dusty Tori, \& Alignment}

Dusty tori can be recognized by their absorbing effects in two ways: first,
by their reddening and obscuration of light at high inclinations, and second,
by dimming direct nuclear light, allowing the detection of a faint, scattered,
polarized spectrum.
The alignment of the polarization of a scattered Type 1 spectrum with the radio
jet is the strongest evidence for a nuclear dusty torus.
Perpendicular or parallel alignment of the scattering polarization was recognized
for a number of Seyfert and radio galaxies by Antonucci
(1982, 1983), and interpreted as arising in the scattering geometries illustrated in
Fig. 4 (see \S4).
The polarimetric evidence has been presented above by giving a few of many examples.

In nearby low luminosity AGN, where high spatial resolution is possible, flattened dusty
regions or their outer 100\,pc scale extensions can be recognized by direct imaging, either 
in absorption or infrared emission (see Fig. 5, and NGC\,1068 above).
Usually the disk axis is well aligned with the radio jets.

There is much evidence for dust in the NLR, and between the NLR and BLR, from 
differential reddening and differential polarization of the broad and narrow lines.
Evidence for flattened dusty disks of high optical depth
in the nuclear regions of low luminosity AGN was noted by de Zotti and
Gaskell (1985).  We note three nice statistical studies relating increasing extinction to
increasing lobe dominance:\\
(i) Hill et al. (1996) sought to
detect the obscured broad line region via observations of Pa$\alpha$ in the near infrared
where reddening effects are much smaller than the optical.  They observed a complete sample
of 3CR FR\,II AGNs, that is, AGN with radio luminosities in the QSO class.
They found increasing reddening with decreasing core-dominance R
(increasing inclination), consistent with a dusty torus model like that shown in Fig.\,2.\\
(ii) Baker and collaborators (1995, 1997) have investigated a complete sample of low-frequency
selected RL QSOs, showing the steepening of the optical continuum and reddening of the Balmer
decrements with decreasing R.\\
(iii) Baker (1997) and earlier, Hes, Fosbury \& Barthel (1994) have shown increasing narrow-line
ratio [O\,II]$\lambda$3727/[O\,III]$\lambda$5007 with decreasing R,
indicating the preferential obscuration of the higher-ionization [O\,III] emission that is
expected to be produced closer to the nucleus.

Alignment of the central engine axis with the host galaxy is important for understanding
Unified Schemes too, and for completeness we
note an investigation by Schmitt et al. (1997) and an earlier
one by Ulvestad and Wilson (1984), showing that the axis of small-scale
radio structure in Seyfert galaxies appeared to avoid alignment with the host galaxy minor axis.
A more recent analysis suggests that this is the result of observational
selection (Nagar \& Wilson 1999).  They find that observations are consistent with a uniform
distribution of radio jet direction relative to the galaxy axis.

Additional important evidence for the existence of dusty tori in all or most luminous AGN is 
the presence
of the 3$\micron$ Bump in essentially all radio-loud QSOs and many radio-quiet QSOs, where the
redshift is low enough to allow detection (\S3.1).
Models suggest that the near-infrared optical depths should depend on inclination (Ward 1995).


\begin{figure}
\plotone{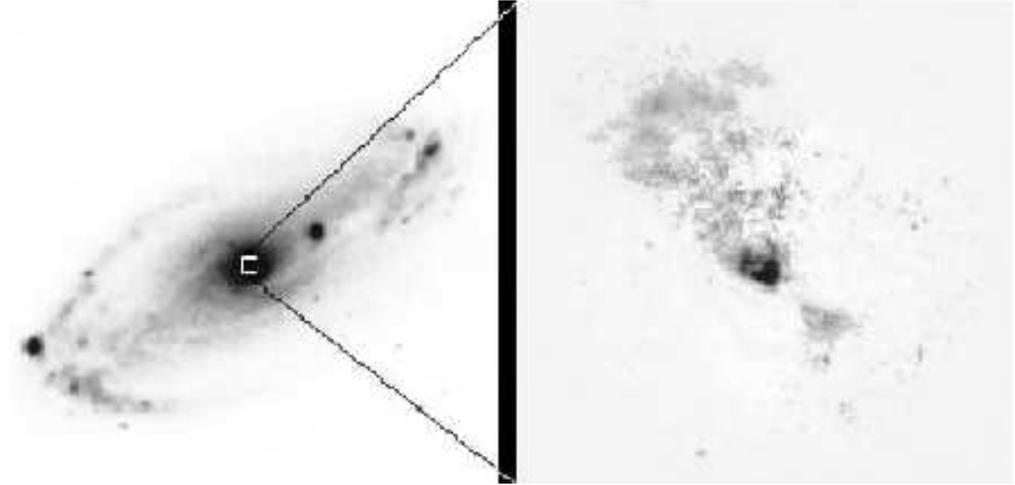}
\caption{The biconical ionization structure in NGC 5728.  An image in the light of the NLR
[O\,III]\,$\lambda\lambda$4959, 5007 emission lines.
Image from http://oposite.stsci.edu/pubinfo/Old.html.  Observation by Wilson et al. (1993).
For other examples, see Wilson \& Tsvetanov (1994).
}
\end{figure}

\subsection{Ionization Cones}

Gas and dust within the opening of the torus, and therefore illuminated by the central QSO,
gives rise to the scattered broad line spectrum, as well as most of the NLR emission.
The structure of these `cones' is clearest for nearby Seyfert 2 galaxies, and more than a dozen
are now known with quite clearly defined `edges'.  However, the emitting gas and dust is obviously clumpy,
so the edges are often ill-defined.  The first ionization cones were 
investigated by Pogge (1989) with arcsecond resolution.  Several examples of bicones are
seen (Fig. 6), but the cone facing away from the observer is often obscured.
Opening angles range from 70\deg\ to 100\deg.  The ionizing continuum deduced from the
ionization state and density of the emitting gas is often
brighter than the observed continuum.  Unified schemes suggest that the dusty torus or
its extensions blocks our view of the continuum.  The dusty torus should be heated by this continuum,
and for a given torus geometry one can predict the amount of re-radiated infrared emission.
Storchi-Bergmann et al. (1992) confirm such predictions for 8 out of a small sample of 9
Seyfert 2s.  A recent survey of early-type Seyfert galaxies (Nagar et al. 1999) and earlier work
show that the axes of ionized emission are typically aligned within a few degrees of the inner 
radio jets.  While some NLR emission is clearly excited by jet-related shocks (e.g., Capetti et al.
1996; Falcke et al. 1998),  
much of the ionized gas is not directly associated with the radio emission;
there is ionized gas beyond the jet and the jets are rather more highly collimated than the ionized
gas.  
Therefore, Wilson (1996) concludes, the radio plasma and ionizing photons are collimated by the same,
or coplanar, disks or tori.

Another test is provided by a comparison of the optical and UV spectra of a sample of Seyfert 2
galaxies.  Kinney et al. (1991) estimate the ionizing photons available from the observed UV continuum,
concluding that in most cases there are insufficient to produce the observed emission line fluxes.
Interestingly, the UV continua, while weak, have slopes indistinguishable from those of Seyfert 1
galaxies, consistent with the Seyfert 2 UV continuum being an electron-scattered  Seyfert 1 continuum.
Comparing a much larger sample of Seyfert 1 and Seyfert 2 AGN, Mulchaey et al. (1994) are able to
show that, while the infrared continua, [O\,III]\,$\lambda$5007 NLR strengths, and the hard X-ray
are similar in both classes, consistent with isotropic emission, the UV and soft X-ray 
(0.2-4 keV) continua of Seyfert 2s are underluminous, consistent with line-of-sight absorption.
In addition, they find that while the line emission and continuum fluxes are correlated
in Seyfert 1s, consistent with the UV photons powering the lines emission, there is no such correlation
for the Seyfert 2s, again consistent with our not seeing the ionizing continuum directly in Seyfert 2s.

There are similar results from calculations of the energy budget for a few higher luminosity 
radio-quiet QSOs [e.g., IRAS\, 09104+4109 (\S4.7); IRAS 20460+1935 (Frogel et al. 1989)].

One can also check the opening angles derived by the above direct imaging results with those
derived by comparing relative numbers of Type 2 and Type 1 AGNs, assuming
that most of the Type 2 AGNs contain buried Type 1 nuclei.  
Opening angles derived from comparisons of Seyfert 1 and Seyfert 2 galaxies
lie in the right ballpark -- about 60\deg\ to 80\deg\ (e.g., Osterbrock \& Shaw 1988).
For RL AGNs, the comparison is easier because it is possible (with a lot of hard work) to select
an orientation-independent sample by isotropic low-frequency radio emission.
To avoid optical bias, the sample must be completely optically identified.  This is how Barthel (1989),
using the completely identified 3CRR sample (178 MHz -- Laing, Riley \& Longair 1983), derived the value
$2\times\sim$45\deg\ used in \S3.1.
Willott, Rawlings \& Blundell (this volume) use the much larger 7C sample selected at 151 MHz, and
derive an opening angle of $\sim$120\deg\ for the most luminous QSOs.  See their paper for a
discussion of this difference.

\subsection{Where are the QSO 2s?}

QSO 2s are to `normal' QSO 1s, as Seyfert 2s are to Seyfert 1s.
It has sometimes been hotly claimed that QSO 2s don't exist, with the suggestion that high luminosity 
AGN have only small covering by a dusty torus.

Investigations of individual QSOs (e.g., \S4) lead to a wide range in dust-covering fraction, and this 
raises the question of just how many luminous QSOs are we missing in samples selected by their optical-UV
emission?
The only way to address this question is to observe a complete sample of AGN selected by a property
that is independent of orientation.  Selection by low-frequency radio emission is the best
method.  The 3CRR sample selected at 178 MHz is excellent, and selection by FR\,II structure identifies
those of QSO luminosities (Urry \& Padovani 1995; Hill et al. 1996).  This was the sample that led
Barthel (1989) to propose that FR\,II radio galaxies were buried FR\,II QSOs.  Thus, some NLRG and
3CR\,68.1 (\S4.5) are QSO 2s, and the BLRG and reddened QSOs (Smith \& Spinrad 1980) are of intermediate
inclination.  See also Willott et al. (this volume).

Recent studies of reddening in RLQs are relevant.
Webster et al. (1995) have found a significant number of red QSOs among a 
completely-identified sample of Parkes flat radio-spectrum sources.  While some fraction 
of these might be expected to have a steep (and therefore red) IR-UV synchrotron component
(Serjeant \& Rawlings 1996, see \S7) a significant fraction are probably dust reddened (Francis
et al. 1997).
Carilli et al. (1998) have found 4 of 5 red flat radio-spectrum QSOs with significant columns
of neutral hydrogen -- either the result of intervening absorbers or, in two cases, intrinsic to the QSO.
This is interesting because, according to unified schemes, these flat-spectrum sources are CD QSOs
and are therefore expected to be the
least dust-reddened, so absorption (and reddening) is likely to be rather more for lobe-dominant
sources where the line-of-sight passes closer to the plane of the dusty torus.

However the above studies apply only to RLQs.

\begin{figure}
\def\plotfiddle#1#2#3#4#5#6#7{\centering
\leavevmode
\vbox to#2{\rule{0pt}{#2}}
\includegraphics{#1}}
\plotfiddle{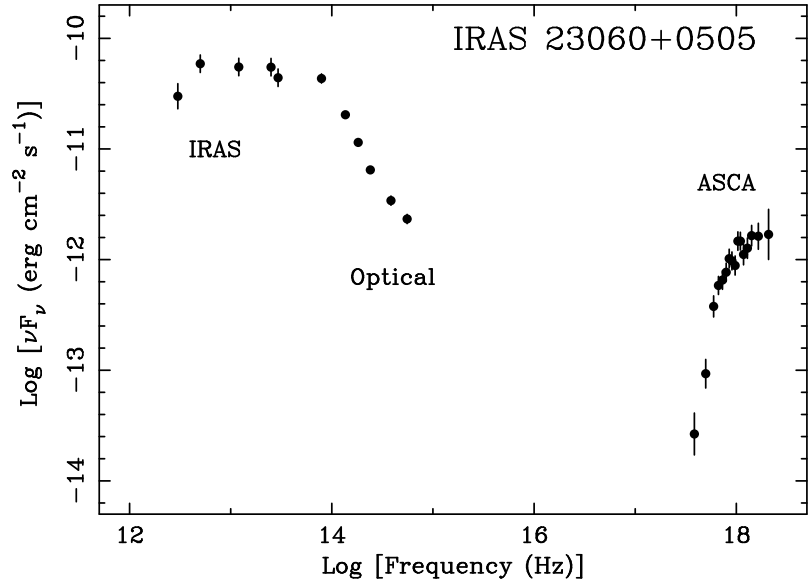}{8.0cm}{-90.}{45.0}{45.0}{-170.}{260.}
\caption{The observed spectral energy distribution of IRAS 23060+0505, a QSO\,2
showing the reddened optical continuum and the X-ray absorption in the ASCA 2--10 keV
band (the decrease in flux density towards soft X-ray energies).
(Figure adapted from Brandt et al. 1997a.)
}
\end{figure}

The RQQs far outnumber the RLQs, so it is of interest to understand how Unified Schemes work
for RQQs.  The hard X-ray region ($>10$keV, Wilkes 1999, this volume) or the mid-infrared
(e.g., Rush, Malkan \& Spinoglio 1993) should be good wavelengths to select a sample because absorption
by cold gas or dust, while not negligible, is much less.  The X-ray data are not yet available.
In the infrared one is overwhelmed by the huge numbers of galaxies, 
whose H\,II regions, excited by young stars, emit thermally in the infrared.  A practical solution
adopted by Low et al. (1988) and others is to select, in addition, by warm infrared colors.  
Neugebauer et al. (1986) had shown that warm infrared colors were an essentially unique signature of QSOs and
Seyferts already discovered by optical techniques, so to find the same kinds of QSOs, independent of 
whether or not their optical-UV emission is covered by dust, Low et al. selected IRAS Type 1 spectra with 
F$_{\nu}(25\mu$m)/F$_{\nu}(60\mu$m) $> 1/4$.  Using the same criteria, Wills \& Hines (1997) defined
a complete sample of extragalactic objects with L(IR)$> 10^{11.8}$L\sun (H$_o = 50$ km s$^{-1}$
Mpc$^{-1}$), regardless of optical classification.  Six objects (1/3 of the sample) had Type 2 spectra
with Type 1 spectra revealed in polarized light, and are therefore QSO 2s.  The spectral energy distribution
of one of these is shown in Fig.\,7, where its steep optical UV spectrum, soft X-ray absorption, 
and recovery at harder X-ray energies, is illustrated.  About 1/3 showed
reddened Type 1 spectra in total and polarized light, providing evidence for a significant 
scattered-light contribution to a partially obscured direct view of the center.  The remainder were 
less reddened, but only 3 of the sample had been found by conventional optical-UV selection techniques.
When corrected for reddening the ratio of IR to optical flux is no different from typical
optically-selected PG QSOs, so spectroscopically and with respect to their spectral energy 
distributions, these IRAS-selected luminous AGN appear to harbor `normal' QSOs.

Two thirds of the sample AGN are either QSO 2s, objects of QSO luminosity dominated by Type 2 spectra
but with buried QSO 1 nuclei, or QSOs without strong NLR emission, but with a reddened line-of-sight
to the center, and a significant scattered Type 1 contribution -- reddened QSO 1s.  These would
not have been discovered by conventional selection techniques.  It is likely that even more-obscured
luminous AGNs were missed by the IRAS color-selection, indicating that space densities of QSOs have been
underestimated by a large factor.

\section{Concluding Remarks}

The present description of Unified Schemes has intentionally been simplified.
Many aspects are the subject of active research. 
There is now a huge literature relevant to the subject -- certainly over 1000 papers.
Therefore I have selected some topics at the expense of others; I apologize for
important papers not included.

Unification of the QSO majority, the most radio quiet, remains difficult without an
indicator of inclination.  One possibility is to search for ionization or scattering cones, 
but these may be small and faint.
One might speculate whether beamed X-ray emission could be used as an orientation indicator
for radio-quiet QSOs.  The basis for this is the discovery of radio-weak BL Lac objects whose
beamed radiation appears to be in the EUV to soft X-ray region (Stocke et al. 1985; Padovani
\& Giommi 1995).  The minimum ingredients for a unification scheme are orientation and patchy
dust, but there is good evidence for a richer and more interesting unified scheme
based on an axisymmetric central engine surrounded by an obscuring dusty torus.

Some tantalizing links are suggested between the torus parameters and other AGN relationships.
Covering by dense Fe\,II-rich gas, and outflowing BAL gas (\S3.3), appear to be part of the
`Principal Component' relationships among X-ray continuum, broad line profiles, and the
inverse relationship between Fe\,II strength and [O\,III]\,$\lambda$5007 (Wills et al.,
Aoki et al., this volume).  Radio loudness appears to be another parameter linked to
these Principal Component relationships and to properties of BAL outflows.
Links between strong Fe\,II emission and dust abound in
astrophysical situations, so, is there a relationship between covering by dense line-emitting
gas and the dusty torus?

It has been suggested that the torus opening angle increases with luminosity.
The X-ray slope and broad-line equivalent widths depend on luminosity (Korista et al.,
this volume).  Broad line widths may also increase with luminosity.

We have yet to learn whether or how all these relationships may be linked.

Unification of diverse classes of AGN by changing aspect angle and extinction by dust
is a reality, as foretold by Rowan-Robinson (1977).  While most astronomers
would like to sweep it under the rug, the complicated subject of dust
must be confronted (e.g., Masci 1998).  Not only does external dust absorb, scatter and emit
thermal radiation, it affects the energy balance and chemistry of absorption and emission 
line regions.  Most surveys for AGN are biased against dusty
objects.  This will change with new surveys in the radio, infrared and hard X-ray regimes.
The lesson for cosmology is that the space density of QSOs has been underestimated,
with important implications for their evolution and contribution to the cosmic background.

With increasing evidence for supermassive black holes in many galaxies, even those with little
activity, it is clear from the space density and luminosity functions of galaxies
and QSOs that QSOs are a phenomenon common to perhaps all massive galaxies, even
if short-lived and intermittent.
Therefore, what we learn about active galaxies and QSOs is relevant to the evolution of all
galaxies, and to cosmology.

\section{Additional References}

Here I expand upon the references given in the above sections, with brief remarks.

\noindent {\bf General AGN Texts:}
Osterbrock 1989, Peterson 1997, Robson 1996, Krolik 1999, Weedman 1986, Longair 1981, Miller 1985,
 Burke \& Graham-Smith 1997.

\noindent {\bf Radio-loud AGN Unification:} 
Blandford \& Rees 1978 (the classic paper on ideas behind present Unified Schemes)
Vermeulen \& Cohen 1994 (superluminal motion)
Ghisellini et al. 1993 (statistics relating radio, optical \& X-ray observations for several classes
of AGN)
Kellermann \& Owen 1988 (general review of extragalactic radio sources at a basic level)
Orr \& Browne 1982 (RL QSO CD vs. LD statistics)
Hughes, P. A. 1991 (Good theory and observation text)
Browne \& Perley 1986 (halos of CD sources)
O'Dea, C. 1998 (Gigahertz Peaked Spectrum sources - a review)
Urry (1999) (The FR\,I--FR\,II break).

\noindent {\bf Unification of FR\,II AGNs:}
Tadhunter, Dickson \& Shaw 1996, Cimatti et al. 1998 (polarimetry of BLRG and reddened QSOs)
Sanders et al. 1989, Barvainis 1987 (3 $\mu$m Bump)
Shastri et al. 1993 (X-ray spectrum and core-dominance)
Wills \& Browne 1986, Vestergaard 1998 (broad-line width vs. core-dominance).

\noindent {\bf FR\,1 AGN \& BL Lac Objects:}
Ulrich et al. 1984, Stickel et al. 1993,
Vermeulen et al. 1995, Corbett et al. 1996 (broad lines in BL Lac objects).

\noindent {\bf Seyfert galaxy unification:}
Tran 1995 (spectropolarimetry),
Veilleux, Goodrich \& Hill 1997 (near IR spectroscopy).
Heisler, Lumsden \& Bailey 1997 (even scattered light obscured at highest inclinations).

\noindent {\bf Radio-quiet QSOs:}
Arav, Shlosman \& Weymann (eds) 1997, Korista et al. 1993, Glenn et al. 1994
     (BAL QSOs, multiple light paths).

\noindent {\bf NGC\,1068:}
Antonucci \& Miller 1985, Bailey et al. 1988, Young et al. 1995 (spectropolarimetry)
Roy et al. 1998 (absorbed radio nucleus)
Bock et al. 1998 (IR imaging)
Capetti et al. 1995 (polarization imaging)
Greenhill, L. 1996 (VLBI H$_2$O maser)
Matt et al. 1997 (hard X-ray spectrum)
Gallimore, Baum, \& O'Dea 1995 (pc-scale thermal emission from the torus).

\noindent {\bf NGC\,4258:}
Miyoshi et al. 1995, Moran et al. 1995 (H$_2$O maser)
Wilkes et al. 1995 (spectropolarimetry).

\noindent {\bf IRAS\,09104+4109:}
Kleinmann et al. 1988 (discovery and spectra)
Hines \& Wills 1993, Hines et al. 1999 (spectropolarimetry and imaging)
Crawford \& Vanderreist 1996 (emission line imaging, X-ray).

\noindent {\bf OI\,287:}
Ulvestad \& Antonucci 1988 (radio structure)
Rudy \& Schmidt 1988, Miller \& Goodrich 1988 (polarization)
Antonucci, Kinney \& Hurt 1993 (UV spectra).

\noindent {\bf 3C\,68.1:}
Brotherton et al. 1998.

\noindent {\bf IRAS\,13349+2438:}
Beichmann et al. 1986 (discovery)
Wills et al. 1992b,  (spectroscopy, polarimetry)
Brandt et al. 1997b (X-ray ionized absorption).

\noindent {\bf IRAS\,07598+6508:}
Lawrence et al. 1988, Lipari 1994, Kwan et al. 1995 (strong Fe\,II emission)
Hines \& Wills 1995 (spectropolarimetry).

\noindent {\bf The dusty torus:}
Lawrence \& Elvis 1982 (perhaps the first paper to suggest that Type 2 spectra
 were Type 1 spectra with absorbed broad line emission and absorbed soft X-rays), 
Ward 1995 (The Oxford Torus Workshop).

\noindent {\bf QSO 2s:}
Fugmann 1988, Impey, Lawrence \& Tapia 1991, Wills 1991, Wills et al. 1992a (IR-optical
synchrotron continuum).

\acknowledgments

I thank Jack Baldwin, Mike Brotherton and Derek Wills for many helpful comments.  I am very
grateful for the work of the staffs of the NASA Astrophysics Data System, the Eprint 
Astrophysics Archives, and the NASA Extragalactic Database, without whom this review would 
have been impossible.
B.J.W. is supported by NASA through LTSA grant number NAG5-3431 and grant number
GO-06781 from the Space Telescope Science Institute, which is operated by
the Association of Universities for Research in Astronomy, Inc., under
NASA contract NAS5-26555.


\begin{references}
\reference Antonucci, R. R. J. 1982 Nature, 299, 605
\reference Antonucci, R. R. J. 1983 Nature, 303, 158
\reference Antonucci, R. R. J. 1993, ARA\&A, 31, 473
\reference Antonucci, R. R. J., \& Miller, J. S. 1985, \apj, 297, 476
\reference Antonucci, R. R. J., Kinney, A. L., \& Hurt, T. 1993, \apj, 414, 506
\reference Antonucci, R. R. J., Hurt, T., \& Miller, J. S. 1994, \apj, 430, 210 
\reference Antonucci, R. R. J., \& Ulvestad, J. S. 1985, \apj, 294, 158
\reference Arav, N., Shlosman, I., \& Weymann, R. A. (eds.) 1997, Mass Ejection from AGN
 (San Francisco: ASP)
\reference Bailey, J. A., Axon, D. J., Hough, J. H., Ward, M. J., McLean, I., Heathcote, S. R. 1988,
 \mnras, 234, 899
\reference Baker, J. C. 1997, \mnras, 286, 23 
\reference Baker, J. C., \& Hunstead, R. W. 1995, \apjl, 452, 95  
\reference Baker, J. C., Hunstead, R. W., \& Brinkmann, W. 1995, \mnras, 277, 553 
\reference Barthel, P. D. 1989, \apj, 336, 606
\reference Barvainis, R. 1987, \apj, 320, 537
\reference Becker, R. H., White, R. L., Helfand, D. J.  1995, \apj, 450, 559
\reference Beichman, C. A., Soifer, B. T., Helou, G., Chester, T. J., Neugebauer, G., Gillett, F. C.,
 \& Low, F. J. 1986, \apjl, 308, 1
\reference Blandford, R. \& Rees, M. 1978, in Pittsburgh Conference on BL Lac Objects, ed.
 A. M. Wolfe (Pittsburgh: University of Pittsburgh), 328
\reference Blundell, K. M., \& Beasley, A. J. 1998a, \mnras, 299, 165
\reference Blundell, K. M., \& Beasley, A. J. 1998b, BAAS, Meeting No. 193, 110.04.
\reference Bock, J. J., Marsh, K. A., Ressler, M. E., Werner, M. W. 1998, \apjl, 504, 5
\reference Boroson, T. A., \& Meyers, K. A. 1992, \apj, 397, 442
\reference Brandt, W. N., Fabian, A. C., Takahashi, K., Fujimoto, R., Yamashita, A., Inoue, H.,
 Ogasaka, Y. 1997a, \mnras, 290, 617
\reference Brandt, W. N., Mathur, S., Reynolds, C. S., \& Elvis, M. 1997b, \mnras, 292, 407
\reference Bridle, A. H., Hough, D. H., Lonsdale, C. J., Burns, J. O., Laing, R. A. 1994, \aj,
 108, 766
\reference Brotherton, M. S., Wills, B. J., Dey, A., Van Breugel, W., \& Antonucci, R. R. J. 1998,
 \apj, 501, 110
\reference Browne, I. W. A. 1987, in Superluminal Radio Sources, eds. J. A. Zensus, \& T. J. Pearson
 (Cambridge: CUP), 129
\reference Browne, I. W. A., \& Perley, R. A. 1986, \mnras, 222, 149
\reference Burke, B. F., \& Graham-Smith, F. 1997, An Introduction to Radio Astronomy 
  (Cambridge: CUP)
\reference Capetti, A., Macchetto, F., Axon, D. J., Sparks, W. B., \& Boksenberg, A. 1995, \apj, 452, L87
\reference Capetti, A., Axon, D. J., Macchetto, F., Sparks, W. B., Boksenberg, A. 1996, \apj, 469, 554
\reference Carilli, C. L., Menten, K. M., Reid, M. J., Rupen, M. P., \& Yun, M. S. 1998, \apj, 494, 175
\reference Chary, R. \& Becklin, E. E. 1997 \apjl, 485, 75
\reference Cimatti, A., Di Serego Alighieri, S., Vernet, J., Cohen, M., \& Fosbury, R. A. E. 1998,
\apj, 499, 21
\reference Cohen, M. H. \& Unwin, S. C. 1982, Proc. IAU Symposium No. 97, ``Extragalactic Radio Sources'',
eds. D. S. Heeschen \& C. M. Wade (Reidel: Dordrecht), 345
\reference Corbett, E. A., Robinson, A., Axon, D. J., Hough, J. H., Jeffries, R. D., Thurston, M. R.,
 Young, S. 1996, \mnras, 281, 737
\reference Crawford, C. S., \& Vanderreist, C. 1996, \mnras, 283, 1003
\reference De Zotti, G., \& Gaskell, C. M. 1985, \aap, 147, 1
\reference Dennett-Thorpe, J., Bridle, A. H., Scheuer, P. A. G., Laing, R. A., Leahy, J. P. 1997,
 \mnras, 289, 753
\reference Falcke, H., Gopal-Krishna, \& Biermann, P. L. 1995, \aap, 298, 395 
\reference Falcke, H., Wilson, A.S., \& Ho, L.C. 1997, Proc. Conf. Relativistic Jets in AGNs, 13
\reference Falcke, H., Patnaik, A. R., Sherwood, W.  1996a, \apjl, 473, 13
\reference Falcke, H., Sherwood, W., Patnaik, A. R. 1996b, \apj, 471, 106
\reference Falcke, H., Wilson, A. S., Simpson, C. 1998, \apj, 502, 199
\reference Fanaroff B. L., \& Riley, J. M. 1974, \mnras, 167, 31P
\reference Ferrarese, L., Ford, H. C., \& Jaffe, W. 1996, \apj, 470, 444
\reference Francis, P.J., Webster, R. L., Masci, F. J., Drinkwater, M. J., Peterson, B. A. 1997,
 in Emission Lines in Active Galaxies: New Methods \& Techniques, ed B. M. Paterson, F.-Z. Cheng,
  \& A. S. Wilson (San Francisco: ASP), 130
\reference Frogel, J. A., Gillett, F. C., Terndrup, D. M., \& Vader, J. P. 1989, \apj, 343, 672
\reference Fugmann, W. 1988, \aap, 205, 86
\reference Gallimore, J. F., Baum, S. A., \& O'Dea, C. P. 1995, Nature, 388, 852
\reference Ghisellini, G., Padovani, P., Celotti, A., Maraschi, L. 1993, \apj, 407, 65
\reference Giovannini, G., Arbizzani, E., Feretti, L., Venturi, T., Cotton, W. D., Lara, L., \&
 Taylor, G. B. 
 1998, IAU Colloquium 164: Radio Emission from Galactic \& Extragalactic Compact Sources, ASP
Conference Ser. 144, eds. Zensus, J.A., Taylor, G.B., \& Wrobel, J.M. (ASP: San Francisco), 85
\reference Glenn, J., Schmidt, G. D., \& Foltz, C. B. 1994, \apjl, 434, 47
\reference Gopal-Krishna, Kulkarni, V. K., Wiita, P. J. 1996, \apjl, 463, 1
\reference Grandi, S. A. \& Osterbrock, D. E. 1978, \apj, 220, 783
\reference Green, P.J., \& Mathur, S. 1996, \apj, 462, 637
\reference Greenhill, L. 1996, \apjl, 472, 21
\reference Heisler, C. A., Lumsden, S. L., \& Bailey, J. A. 1997, Nature, 385, 700
\reference Hes, R., Fosbury, R. A. E., \& Barthel, P. D. 1994, in The First Stromlo Symposium:
The Physics of Active Galaxies, ASP Conference Ser. 54, eds. G. V. Bicknell, M. A. Dopita,
\& P. J. Quinn (ASP: San Francisco), 367
\reference Hill, G. J., Goodrich, R. W., \& DePoy, D. L. 1996, \apj, 462, 163
\reference Hines, D. C., Schmidt, G., Wills, B. J., Smith, P. S. \& Sowinski, G. 1999, \apj, 512, 000
\reference Hines, D. C., \& Wills, B. J. 1993, \apj, 415, 82
\reference Hines, D. C., \& Wills, B. J. 1995, \apjl, 448, 69
\reference Hooper, E. J., Impey, C. D., Foltz, C. B., Hewett, P. C.  1995, \apj, 445, 62
\reference Hough, D. H., \& Readhead, A. C. S. 1989, \aj, 98, 1208
\reference Hughes, P. A. (ed.) 1991, Beams \& Jets in Astrophysics (Cambridge: CUP)
\reference Impey, C. D., Lawrence, C. R., \& Tapia, S. 1991, \apj, 375, 461
\reference Jaffe, W., Ford, H. C., Ferrarese, L., van den Bosch, F., \& O'Connell, R. W. 1993, 
 Nature, 364, 213 
\reference --- 1996, ApJ, 460, 214
\reference Jones, D.L. \& Wehrle, A. E. 1998,
IAU Colloquium 164: Radio Emission from Galactic \& Extragalactic Compact Sources, ASP
Conference Ser. 144, eds. Zensus, J.A., Taylor, G.B., \& Wrobel, J.M. (ASP: San Francisco), 63
\reference Kellermann, K. I., \& Owen, F. N. 1988, in Galactic \& Extragalactic Radio Astronomy,
 second edition,
 eds. K. I. Kellermann, \& G. L. Verschuur (New York: Springer-Verlag), 563
\reference Kellermann, K. I., Sramek, R. A., Schmidt, M., Green, R. F., Shaffer, D. B. 1994, 
\aj, 108, 1163
\reference Kellermann, K. I., Vermeulen, R. C., Zensus, J. A., \& Cohen, M. H. 1998,
 \aj, 115, 1295
\reference Kinney, A. L., Antonucci, R. R. J., Ward, M. J., Wilson, A. S., Whittle, M. 1991,
 \apj, 377, 100
\reference Kleinmann, S. G., Hamilton, D., Keel, W. C., Wynn-Williams, C. G., Eales, S. A.,
 Becklin, E. E., Kuntz, K. D. 1988, \apj, 328, 161
\reference Korista, K. T., Voit, G. M., Morris, S. L., \& Weymann, R. J. 1993, \apjs, 88,357
\reference Krolik, J. H. 1999, Active Galactic Nuclei (Princeton: Princeton University Press)
\reference Kukula, M. J., Dunlop, J. S., Hughes, D. H., \& Rawlings, S. 1998, \mnras, 297, 366
\reference Kwan, J., Cheng, F.-Z., Fang, L.-Z., Zheng, W., Ge, J. 1995, \apj, 440, 628
\reference Laing, R.A., Riley, J.M., \& Longair, M.S. 1983, \mnras, 204, 151
\reference Lara, L., M\'arquez, I, Cotton, W. D., Feretti, L., Giovannini, G., 
  Marcaide, J. M., \& Venturi, T. 1998, astro-ph/9812254
\reference Lawrence, A., \& Elvis, M. 1982, \apj, 256, 410
\reference Lawrence, A. et al. 1988, \mnras, 235, 261
\reference Leahy, J. P., et al. 1997, \mnras, 291, 20
\reference Lind, K. R., \& Blandford, R. D. 1985, \apj, 295, 358
\reference Lipari, S. 1994, \apj, 436, 102
\reference Longair, M. S. 1981, High Energy Astrophysics (Cambridge: CUP)
\reference Low, F. J.,Cutri, R. M., Huchra, J. P., Kleinmann, S. G. 1988, \apjl, 327, 41
\reference Malkan, M. A. \& Sargent, W. L. W. 1983, \apj, 254, 22
\reference Masci, F. J. 1998, PASA, 15, 299
\reference Matt, G., et al. 1997, \aap, 325, L13
\reference Miller, J. S., \& Antonucci, R. R. J. 1983, \apjl, 271, 7
\reference Miller, J. S., \& Goodrich, R. W. 1988, \apj, 331, 332
\reference Miller, J. S., Goodrich, R. W., \& Mathews, W. G. 1991, \apj, 378, 47
\reference Miyoshi, M., Moran, J. M., Herrnstein, J. R., Greenhill, L. J., Nakai, N., 
 Diamond, P. J., \& Inoue, M. 1995, Nature, 373, 127
\reference Morganti, R., Parma, P., Capetti, A., Fanti, R., De Ruiter, H.R., 1997, \aap, 326, 919
\reference Mulchaey, J. S., et al. 1994, \apj, 436, 586
\reference Mushotzky, R. F. 1997, in Mass Ejection from AGN, ASP
  Conference Ser. 128, eds. N. Arav, I. Shlosman, \& R. J. Weymann (ASP: San Francisco), 141
\reference Moran, J. M., Greenhill, L. J., Herrnstein, J. R., Diamond, P. J., Miyoshi, M., 
 Nakai, N., \& Inoue, M. 1995, Proc. Natl. Acad.  Sci., 92, 11427
\reference Morganti, R., Parma, P., Capetti, A., Fanti, R. \& de Ruiter, H. R. 1997, \aa, 326, 919
\reference Nagar, N. M., \& Wilson, A. S. 1999, \apj, 516, No. 1
\reference Nagar, N. M., Wilson, A. S., Mulchaey, J. S., \& Gallimore, J. F. 1999, \apjs, 120, No. 2
\reference Neeser, M. J., Eales, S. A., Law-Green, J. D., Leahy, J. P., Rawlings, S. 1995, \apj, 451, 76
\reference Neugebauer, G., Soifer, B. T., Miley, G. K., \& Clegg, P.E. 1986, \apj, 308, 815
\reference O'Dea, C. P. 1998, PASP, 110, 493
\reference Orr, M. J. L., \& Browne, I. W. A. 1982, \mnras, 200, 1067
\reference Osterbrock, D. E. 1989, Astrophysics of Gaseous Nebulae \& Active Galactic Nuclei
  (Mill Valley: University Science Books)
\reference Osterbrock, D. E., \& Shaw, R. A. 1988, \apj, 327, 89
\reference Owen, F. N., Ledlow, M. J., \& Keel, W. C. 1996, \aj, 111, 53
\reference Packham, C., Young, S., Hough, J. H., Axon, D. J., Bailey, J. A. 1997, \mnras, 288, 375
\reference Padovani, P., \& Giommi, P. 1995, \apj, 444, 567
\reference Penston, M. V., \& Perez, E. 1984, \mnras, 211, 33P
\reference Peterson, B. M. 1997, An Introduction to Active Galactic Nuclei (Cambridge:CUP)
\reference Pogge, R. W. 1988, \apj, 328, 519
\reference Pogge, R. W. 1989, \apj, 345, 730
\reference Robson, I. Active Galactic Nuclei (Chichester: Wiley-Praxis)
\reference Roy, A. L., Colbert, E. J. M., Wilson, A. S., Ulvestad, J. S. 1998, \apj, 504, 147
\reference Rudy, R. J. \& Schmidt, G. D. 1988, \apj, 331, 325
\reference Rush, B., Malkan, M. A., \& Spinoglio, L. 1993, \apjs, 89, 1
\reference Sanders, D. B., Phinney, E. S., Neugebauer, G., Soifer, B. T., Matthews, K. 1989,
 \apj, 347, 29
\reference Schmitt, H. R., Kinney, A. L., Storchi-Bergmann, T., \& Antonucci, R. R.
1997, \apj, 477, 623
\reference Shastri, P., Wilkes, B. J., Elvis, M., McDowell, J.  1993, \apj, 410, 29
\reference Shields, G. A. 1978, Nature, 272, 706
\reference Smith, H. E., \& Spinrad, H. 1980, \apj, 236, 419
\reference Stickel, M, Fried, J. W., \& K\"uhr, H. 1993, \aaps, 98, 393
\reference Stocke, J. T., Liebert, J., Schmidt, G, Gioia, I. M., Maccacaro, T., Schild, R. E.,
 Maccagni, D., \& Arp, H. C. 1985, \apj, 298, 619
\reference Storchi-Bergmann, T., Mulchaey, J. S., \& Wilson, A. S. 1992, \apj, 395, 73
\reference Tadhunter, C. N., Dickson, R. C. \& Shaw, M. A. 1996 \mnras, 281, 591
\reference Tran, H. D. 1995, \apj, 440, 597
\reference Tran, H. D., Brotherton, M. S., Stanford, S. A., van Breugel, W., Dey, A., Stern, D., \&
 Antonucci, R. R. J. 1999, \apj, in press
\reference Trotter, A. S., Greenhill, L. J., Moran, J. M., Reid, M. J., Irwin, J. A., \& Lo, K.-Y.
 1998, \apj, 495, 740
\reference Turnshek, D. A., Monier, E. M., Sirola, C. J., \& Espey, B. R. 1997, \apj, 476, 40
\reference Ulrich, M. H., Hackney, K. R. H., Hackney, R. L., \& Kondo, Y. 1984, \apj, 276, 466
\reference Ulvestad, J. S., \& Antonucci, R. R. J. 1988, \apj, 328, 569
\reference Ulvestad, J. S., \& Wilson, A. S. 1984, \apj, 285, 439
\reference Urry, C. M. 1999, in The BL Lac Phenomenon (Turku, Finland)
\reference Urry, C. M., \& Padovani, P. 1995, PASP, 107, 803
\reference Veilleux, S., Goodrich, R. W., Hill, G. J. 1997, \apj, 477, 631
\reference Vermeulen, R. C., \& Cohen, M. H. 1994, \apj, 430, 467
\reference Vermeulen, R. C., et al. 1995, \apjl, 452, 15
\reference Vestergaard, M. 1998, PhD thesis, Univ. of Copenhagen
\reference Ward, M. J. (ed.) 1994, Oxford Astrophysics Workshop on Evidence for the Torus
\reference Wardle, J. F. C., \& Aaron, S. E. 1997, \mnras, 286, 425
\reference Webster, R. L., Francis, P. J., Peterson, B. A., Drinkwater, M. J., \& Masci, F. J. 1995,
 Nature, 375, 469
\reference Weedman, D. W. 1986, Quasar Astronomy (Cambridge: CUP)
\reference Weymann, R. J., Morris, S. L., Foltz, C. B., Hewett, P. C. 1991, \apj, 373, 23
\reference Weymann, R. J. 1997, in Mass Ejection from AGN, ASP
  Conference Ser. 128, eds. N. Arav, I. Shlosman, \& R. J. Weymann (ASP: San Francisco), 3
\reference Wilkes, B. J., Schmidt, G. D., Smith, P. S., Mathur, S., \& McLeod, K. K. 1995, \apjl, 455, L13 
\reference Wills, B. J. 1991, in Variability of Active Galactic Nuclei, eds. H. R. Miller \&
P. J. Wiita (Cambridge: CUP), 87
\reference Wills, B. J., \& Browne, I. W. A. 1996 \apj, 302, 56
\reference Wills, B. J., \& Hines, D. C., in Mass Ejection from AGN, ASP
  Conference Ser. 128, eds. N. Arav, I. Shlosman, \& R. J. Weymann (ASP: San Francisco), 99
\reference Wills, B. J., Wills, D., Breger, M., Antonucci, R. R. J., \& Barvainis, R. 1992a, \apj, 398, 454
\reference Wills, B. J., Wills, D., Evans, N. J., Natta, A., Thompson, K. L., \& Breger, M. 1992b, \apj,
 400, 96
\reference Wilson, A. S. 1996, Vistas in Astronomy, 40, 63
\reference Wilson, A. S., \& Tsvetanov, Z. I. 1994, \aj, 107, 1227
\reference Wilson, A. S., Braatz, J. A., Heckman, T. M., Krolik, J. H., \& Miley, G. K. 1993, \apjl, 419, 61
\reference White, R. L. et al. 1999, \aj, in press
\reference Young, S., Hough, J. H., Axon, D. J., Bailey, J. A., Ward, M. J. 1995, \mnras, 272, 513

\end{references}
\end{document}